\documentstyle[prl,aps,twocolumn]{revtex}
\tightenlines
\begin{document}
\twocolumn[\hsize\textwidth\columnwidth\hsize\csname
@twocolumnfalse\endcsname
\draft
\title{Classical Communication Cost in Distributed Quantum
Information Processing---A generalization of Quantum Communication
Complexity}
\author{Hoi-Kwong Lo}
\address{MagiQ Technologies Inc.\\
275 Seventh Avenue\\
26th Floor\\
New York, NY 10001}

\date{\today}
\maketitle
\begin{abstract}

We study the amount of classical communication needed for distributed
quantum information processing. In particular, we introduce the
concept of "remote preparation" of a quantum state. Given an ensemble of
states, Alice's task is to help Bob in a distant laboratory to prepare
a state of her choice. We find several examples of an ensemble with an
entropy $S$ where the remote preparation can be done with a communication
cost lower than the amount ($2S$) required by standard teleportation.
We conjecture that, for an {\it arbitrary} $N$-dimensional
{\it pure} state, its remote preparation requires $2{ \log_2} N$
bits of classical communication, as in standard teleportation.

\end{abstract}
\pacs{PACS Numbers:}
]
\narrowtext
\section{Introduction and Motivation}
\label{Intro}
There are two main motivations for studying the classical
communication cost in quantum information processing (CCCIQIP).
The first motivation is to better understand the fundamental laws of
quantum information processing. The second is the fact that
CCCIQIP can be regarded as a natural generalization of quantum
communication complexity, a subject of much recent interest.

\subsection{First Motivation}
Quantum information theory---the
synthesis of quantum mechanics with information theory---has
been a subject of much recent interest. It is now known
that novel phenomena including teleportation \cite{tele} and
dense coding \cite{dense}
can occur when the laws of quantum mechanics are invoked in
information processing. To better understand those diverse
exotic phenomena, it is important to derive the fundamental
laws of quantum information processing.

Until recently, it was customary to ignore the classical
communication cost in quantum information processing.
The motivation was that classical communication is ``cheap''
whereas quantum communication and entanglement are expensive.
However, as emphasized in \cite{LP99}, in applications such
as dense coding\cite{dense}, classical
communication cost is of primary interest
and it would be totally inconsistent to ignore it.
In summary, it is important to take full consideration of
classical communication cost in the study of quantum information
processing.

Some examples of CCCIQIP (for example, ``remote preparation'' to be
introduced in this paper) can also be regarded
as a refinement of Schumacher's coding theorem\cite{Sch} in
which information is decomposed into two parts:
(a) a quantum piece (prior entanglement) and (b) a classical piece
(subsequent classical communication).
In contrast, in the standard Schumacher's coding theorem, quantum
information is transmitted directly via quantum bits (qubits).

\subsection{Second Motivation}
In contrast to the lack of interest in classical communication cost
shown in the quantum community, classical communication cost
is an important subject in theoretical computer science. It is given the name
''communication complexity''.
For example, two or more parties with distributed private
inputs $i_1, i_2, \cdots, i_N$ would like to cooperate to compute a function
$f(i_1, i_2, \cdots, i_N)$. (For instance, in an appointment scheduling
problem,
two distant parties would like to find a date when both are free.)
They do so by sending classical bits to each other.
The goal of communication complexity is to study the number of classical
bits of
communication needed.
Of particular interest is the limiting case when the problem ''size'' is big.
(For instance,
in the appointment scheduling
problem, the number of dates under consideration is large.)
Classical communication complexity can be regarded as the study of
classical communication resource (classical bits) in a classical problem.

Recently, there has been much interest in using {\it quantum} resources,
namely prior
entanglement, to reduce the communication complexity of a classical function.
While some problems are now known to allow a huge reduction,
problems such as the inner product function have been shown to forbid any
saving. Quantum communication complexity \cite{commcomp}
can, therefore, be regarded as the study of entanglement-enhanced
communication complexity of a {\it classical} function.

In quantum information processing, a new complication arises:
the input and output states may be non-classical.
The simplest example is an entangled state.
(More subtle form of nonlocality without entanglement
also exists \cite{nonlocality}.)
The study of classical communication cost in quantum
information processing can, thus, be regarded as a natural
generalization of quantum communication complexity by allowing
the inputs and outputs to be (possibly non-separable) quantum states,
rather than
classical ones.

\subsection{Prior Works}
There are a number of prior works. The first paper on the subject of
CCCIQIP is probably the seminal teleportation paper\cite{tele},
in which it is
shown that an arbitrary unknown state (possibly entangled with an external
system) in an $N$-dimensional Hilbert space can be transmitted
by the dual usage of prior entanglement and $2 \log_2 N$
classical bits of communication.\footnote{The amount of classical
communication,
$2 \log_2 N$ bits, needed is optimal. This follows from dense coding
\cite{dense}.
If transmission of an arbitrary (possibly entangled)
$N$-dimensional state could be done with less than $2 \log_2 N$ bits
of classical communications, then causality would be violated.}

Notice that, in the classical case, if the value of the input $( i, j)$
and the
deterministic output
$f(i,j)$ are fixed and given beforehand, the classical communication complexity is
trivially zero. The quantum case (CCCIQIP) is strikingly different.
Even if Alice and Bob know exactly their fixed input $\Psi$ and output $\Phi$
states, the manipulation of a bi-partite state $\Psi$  into another bi-partite $\Phi$
state may still require a non-trivial amount of classical
communication\footnote{This result
is not difficult to prove, using the idea of the proof in
\cite{LP97} that
entanglement manipulation strategies with one-way
communication is generally more
powerful than those with no communication.}
The intuitive reason behind this result is that a quantum state is
generally entangled.
Since it cannot be written as a direct product of pure states, it cannot be
prepared by bi-local operations. See also  \cite{nonlocality}.

Interestingly, in some situation the classical communication cost can be made to
vanish in the
asymptotic limit \cite{LP99}. Consider the situation of entanglement dilution \cite{BBPS}:
two distance observers Alice and Bob who share a large number $NS$ singlet
 (i.e., maximally entangled
states) would like to dilute them into
$N$ pairs of the non-maximally entangled state $ a | 00 \rangle + b | 11 \rangle$ whose
entropy of entanglement, $entropy  = - |a|^2 \log_2 |a|^2 - |b|^2
 \log_2| b|^2 $  is
equal to $S$.
The standard scheme \cite{BBPS} involves a teleportation step and, thus, has a
classical communication cost proportional to $N$.
Nevertheless, it was subsequently shown in \cite{LP99}
that entanglement dilution
can be done in the asymptotic
limit with a vanishing amount of classical communication.
As a consequence, entanglement is, indeed, a fungible resource.
That is to say that the same amount of two-party pure
state entanglement in different forms or concentrations
can truly be regarded as equivalent because they are interconvertible
into each other \cite{BBPS}, with a
negligible amount of classical communication
cost between the two parties \cite{LP99}.
A key point of their argument is that there is a huge
degree of degeneracy in the Schmidt coefficients \cite{Appen}
of the relevant bi-partite state.

Recently, important discussions on the classical communication cost of
entanglement manipulations
has also been made by Nielsen \cite{Nielsen9909}

\subsection{Related Works}
CCCIQIP is also related to other subjects. For instance, Brassard,
Cleve and Tapp \cite{BCT}
have studied the issue of simulating entanglement with
classical communication. Another related subject is quantum non-locality
without entanglement \cite{nonlocality}. It concerns the opposite question,
namely the crucial role of
quantum entanglement in a rather novel context.

CCCIQIP and many other studies can be regarded as the investigations of
limiting cases of quantum information processing, in which the cost of
one type of resources (entanglement or classical communication) is often ignored.

\subsection{Main Result}
Our main results are as follows:
First of all, as first pointed out by
Daniel Gottesman \cite{got}, the usual
teleportation \cite{tele} can be decomposed into a two-stage process.
Starting with a pure state $a | 0 \rangle + b | 1 \rangle$ in
Alice's side and an EPR pair shared between
Alice and Bob, the first stage will lead to an entangled
state $a| 00 \rangle + b | 11 \rangle$ shared between
Alice and Bob. The second stage will lead to a state
$a| 0 \rangle + b | 1 \rangle$ fully in Bob's hand.
Moreover, each stage requires a single bit of classical
communication. (This works not only for a pure
initial state, but also for a qubit that is
entangled with an ancilla.)

Second, we give a simple procedure that halves the amount of
classical communication cost in entanglement dilution
compared to even the improved scheme in \cite{LP99}. This is done by
noting that only the first stage of
teleportation is needed for entanglement dilution.
(The second step can be simply skipped.)

Third, we move on to consider the
following general problem, which we shall call ``remote
preparation''  following Popescu\cite{pop1}.
Suppose Alice and Bob initially share some entanglement.
We only allow Alice to send classical bits to Bob.
Alice's
goal is to help Bob to prepare some pure state chosen from some
specific pre-agreed distribution. The key difference
between remote preparation and the usual
teleporation is that, unlike teleportation,
we assume in remote preparation
that Alice knows the precise state of
the object that she is trying to help Bob to prepare.
(Put it differently, Alice is given an infinite number of
copies of the pure state and is required to transmit only one to
Bob.)
In addition, the pre-agreed distribution may not be
totally random.
We have some theorems and a conjecture.
With some appropriate constraints on the
distribution, one can use our theorems to
reduce the classical communication cost
below what is required in teleportation even in the asymptotic case.
In other words, remote preparation of constrained states
in some cases offers a discount rate compared to
full-blown teleportation of an ensemble with the same
amount of entropy. This is {\it a priori} surprising result.
We conjecture that
such reduction in classical communication cost is impossible for
an unconstrained state.

\section{Two stage teleportation}
Suppose Alice would like to transmit an unknown qubit
$a | 0 \rangle_q + b | 1 \rangle_q $ to Bob.
Instead of sending it directly to Bob via a quantum
communication channel, Alice can achieve
the same goal by using a classical channel, provided that
Alice and Bob initially share some entanglement.
This process is called teleportation \cite{tele}. Transmission of each
qubit requires two classical bits of communication.
(It can be shown that teleportation works not only for
pure states, but also for states that are entangled
with ancillas.)
In what follows, the well-known teleportation process will be
decomposed
into two steps. The following result was pointed out by
Gottesman \cite{got}.

{\bf Theorem~1: Two-stage teleportation.} Suppose Alice and Bob share
an EPR pair and that Alice is given an unknown qubit
$a | 0 \rangle_q + b | 1 \rangle_q $ in her hand.
There exists a two-stage process for transmitting the
unknown qubit to Bob such that,
on completion of the first
step, Alice shares with Bob an entangled
state $a| 00 \rangle_{AB} + b | 11 \rangle_{AB}$ and
on completion of the second step,
the state
$a| 0 \rangle_B + b | 1 \rangle_B $ is fully in Bob's hand.
Furthermore,
each step requires a single bit of classical
communication.

{\it Remark~A}: Essentially the same procedure works for
an initial state that is entangled with an ancilla.

{\it Remark~B}: An analogous procedure works for $N$-dimenssional
state with $\log_2 N$ classical bits of
communication needed for each step.

{\bf Proof}: {\it Step 1}: Alice applies an exclusive
OR (XOR) between the unknown qubit, $q$, and her member,
$A$, of the EPR pair
that she shares with Bob, with the unknown qubit as the target qubit.
Since
\begin{eqnarray}
| a \rangle_q | 0 \rangle_A & \to & | a \rangle_q | 0 \rangle_A
\nonumber \\
| a \rangle_q | 1 \rangle_A & \to & | a+ 1 \rangle_q | 1 \rangle_A
,
\end{eqnarray}
one gets
\begin{eqnarray}
& & a | 0 \rangle_q + b | 1 \rangle_q
\left( | 00 \rangle_{AB} + | 11 \rangle_{AB} \right) \nonumber \\
& \to & | 0 \rangle_q \left( a| 00 \rangle_{AB} + b | 11 \rangle_{AB}
\right) \nonumber \\
& & + | 1 \rangle_q \left( b| 00 \rangle_{AB} + a | 11 \rangle_{AB}
\right) .
\end{eqnarray}

Now, Alice measures the qubit $q$ and sends the outcome, a single
bit, via a classical communication channel to Bob.
If the outcome is $0$, Alice and Bob share
$ a| 00 \rangle_{AB} + b | 11 \rangle_{AB}$ as required.
If the outcome is $1$, they share
$b| 00 \rangle_{AB} + a | 11 \rangle_{AB}$.
Alice and Bob can now apply a bi-local unitary transformation
$| 0 \rangle \to | 1 \rangle$ to obtain the desired state
$a| 00 \rangle_{AB} + b | 11 \rangle_{AB}$.

{\it Step~2}: Alice applies a Hadamard transformation on her
member of the shared pair. She then measures it and sends the
outcome to Bob. On receiving Alice's outcome, Bob applies a
unitary transformation on his member of the shared pair to
recover the unknown qubit.
Mathematically, the Hadamard transform is, up to an overall
normalization,
\begin{eqnarray}
| 0 \rangle_ A & \to & | 0 \rangle_A + | 1 \rangle_A \nonumber \\
| 1 \rangle_ A & \to & | 0 \rangle_A - | 1 \rangle_A .
\end{eqnarray}

Therefore,
\begin{eqnarray}
& & a| 00 \rangle_{AB} + b | 11 \rangle_{AB} \nonumber \\
& \to & a \left( | 0 \rangle_A + | 1 \rangle_A \right) | 0 \rangle_B
+ b \left( | 0 \rangle_A - | 1 \rangle_A \right) | 1 \rangle_B
\nonumber \\
&= & | 0 \rangle_A \left( a | 0 \rangle_B + b | 1 \rangle_B \right)
+ | 1 \rangle_A \left( a | 0 \rangle_B - b | 1 \rangle_B
\right).
\end{eqnarray}
Now Alice measures $A$. If she obtains $0$ as the outcome,
then Bob has $a | 0 \rangle_B + b | 1 \rangle_B $ as required.
Similarly, if Alice obtains $1$ as the outcome,
Bob then has $a | 0 \rangle_B - b | 1 \rangle_B$ which
can now be converted to $a | 0 \rangle_B + b | 1 \rangle_B $
by applying the Pauli operator, $\sigma_z$.~QED.

For experts in stablizer codes, the above result is rather trivial.
However,  theorem~1 has a simple application on entanglement dilution:

{\bf Corollary~2}: One can halve the amount of classical
communication needed for entanglement dilution.

{\bf Proof}: For entanglement dilution, the desired output
state of Bob is entangled with Alice. Therefore, all is
required is the first step of the two-step teleportation
procedure. By skipping the second step, one saves half of the
classical communication cost.

{\it Remark~C}: Corollary~2 applies not only to a naive entanglement
dilution scheme, but also to the advanced scheme proposed
in \cite{LP99}, which requires an vanishing amount of
classical communication in the asymptotic limit.

\section{Remote Preparation of Constrained States}
So far our discussion has been restricted to
teleportation.
It turns out that the
idea of decomposing the transmission process of quantum
information into two parts, as employed in Theorem~1,
is useful in a more general context.
In this Section, we illustrate this point by considering
a similar but more general procedure for transmitting quantum
information, which has been called ''remote preparation'' by Popescu.
Suppose Alice and Bob initially share some entanglement
and subsequently Alice can only send classical bits to Bob.
The goal of remote preparation is
for Alice to help Bob to prepare some pure state chosen from some
specific pre-agreed distribution. The big difference
between remote preparation and the usual
teleporation is that, unlike teleportation,
in remote preparation, Alice knows the precise pure state of
the object that she is trying to help Bob to prepare.
(Equivalently, Alice is given an infinite number of
copies of the pure state and is required to transmit only one to
Bob.) Another difference is that, in general,
the pre-agreed distribution does not need to be random.
We have the following asymptotic (large $N$) result.

{\bf Theorem~3}: Suppose Alice and Bob
are given the values of $a$ and $b$ and that
they satisfy $| a|^2 + |b|^2 =1$.
Suppose further that Alice and Bob share $NS$ ebits of entanglement,
where $S = - | a|^2 \log_2 | a|^2 - |b|^2 \log_2 | b|^2$
for a large $N$.
Alice would like to help Bob to remotely prepare
$N$ objects, each of the form
$a| 0 \rangle + b e^{i \theta_i} | 1 \rangle$.
Here $\theta_i$'s are known to Alice only.
We claim that $NS$ bits of classical communication is
sufficient.

{\bf Proof}. By entanglement (concentration and) dilution
\cite{BBPS,LP99}, Alice and
Bob can convert the $NS$ ebits of entanglement into
$N$ pairs of $a | 0 0 \rangle + b | 11 \rangle$ with
a very high fidelity (and with asymptotically vanishing
amount of classical communication \cite{LP99}).
Now, consider a two stage remote preparation process
in complete analogy with two-stage teleporation.
(i.e., from $a| 0 \rangle_a + b e^{i \theta_i} | 1 \rangle_a $
to $a| 00 \rangle_{AB} + b e^{i \theta_i} | 11 \rangle_{AB}$ and
then $a| 0 \rangle_B + b e^{i \theta_i} | 1 \rangle_B $.)
Since Alice and Bob already share
$a | 0 0 \rangle + b | 11 \rangle$,
they now have a short-cut to step~1.
Indeed, they can convert $a | 0 0 \rangle + b | 11 \rangle$ into
$ a| 00 \rangle + b e^{i \theta_i} | 11 \rangle$, for each $i$,
with no communication at all: Using her knowledge of $ \theta_i$,
this can be done by Alice's rotating
the phase of $| 1 \rangle$ under her control,
i.e., $| 1 \rangle \to e^{i \theta_i} | 1 \rangle$.
Now each of
Alice and Bob
then performs quantum data compression \cite{Sch} on his/her
system, compressing it into $NS$ qubits. (Notice that $S$ is the von
Neumann
entropy of Alice/Bob's system.)
Alice and Bob then perform the
second step of the two-step teleportation process---more precisely,
its higher dimensional generalization as noted in Remark~B---
on the typical space, which has a dimension $2^{O(NS)}$.
This requires $NS$ classical bits and zero e-bits.
Bob can now perform quantum date dilution to
recover the system.~QED.

{\it Remark~D}: As far as classical communication
cost is concerned, our result is optimal.
That is to say that $NS$ bits are necessary for the remote
preparation of the above ensemble. The reason is the following.
$S$ is the von Neumann entropy of Alice's ensemble (i.e.,
a random ensemble of pure state of the form $a| 0 \rangle_B + b e^{i \theta_i}
 | 1 \rangle_B $).
By Holevo's Theorem \cite{Holevo}, the quantum signals can be used to
transmit $NS$ classical bits to Bob. So, if there
were a way to
transmit the quantum signals to Bob with fewer than
$NS$ bits of classical bits, causality would be violated.

{\it Remark~E}: The special case where $|a| = |b|$ has also been proven
by various people including Popescu \cite{pop}.

So far, our discussion has focussed on the classical communication cost.
What about the amount of quantum resource (entanglement) for remote
preparation?
We have the following conjecture.

{\bf Conjecture~4}: For any remote preparation procedure that
uses only $NS$ bits, $NS$ ebits is the minimal amount
of entanglement needed for any remote preparation procedure for the
$N$ signals of the form
$a| 0 \rangle + b e^{i \theta_i} | 1 \rangle$ where $a$ and $b$ are known to
Alice and Bob and $\theta_i$'s are known to Alice only, and
$S = - | a|^2 \log_2 | a|^2 - |b|^2 \log_2 | b|^2$.

We remark that the proviso---that uses only $NS$ bits---is necessary.
Without such a proviso, less ebits can be used at the expense of
a large number of bits. For example, Alice can divide the
latitude into two semi-circular segments, $ 0 \leq \theta \leq \pi$
and $\pi \leq \theta  \leq 2 \pi$. Now, she encodes a state by
two pieces, one classical and one quantum. For a state in
the first segment, she encodes it
by a classical bit $0$, together with the state as it is.
For a state in the second segment, she encodes it by a classical bit
$1$ together with the rotated
$a| 0 \rangle + b e^{i ( \theta_i- \pi)} | 1 \rangle $,
which is a state in the first segment.
Notice that, the quantum states, now all being in the same
segment, have a smaller entropy than $NS$ e-bits.
A similar reasoning can be used to reduce the e-bit cost in
remote preparation even further at the expense of
increasing classical bit cost.

\section{Further examples of Remote Preparation}
In Theorem~3, the moduli, $a$ and $b$,
of the coefficients of each state are independent of $i$.
One might wonder if this is a necessary condition for the reduction
in classical communication cost. The answer is no.
Indeed, in the following theorem, Theorem~5, we give
an example in which reduction of classical communication
cost happens even when the moduli of the coefficients
vary for the states in the ensemble.
What is actually needed is some constraint on
those coefficients.

{\bf Theorem~5}: Suppose Alice and Bob initially share
entanglement and only
a classical communication channel.
Alice would like to help Bob to
prepare a set of $N$ normalized states, where each state,
say the $i$-th one, is of the form,
$a_i | 0 \rangle + b_i | 1 \rangle + c_i | 2 \rangle + d_i| 3 \rangle$
with
\begin{equation}
| a_i |^2 + | b_i |^2 = 2e^2
\label{eq:constraint}
\end{equation}
for all $i$'s.
Here, $e$ is known in advance to both Alice and Bob
whereas only Alice knows the individual coefficients
$a_i$, $b_i$, $c_i$ and $ d_i$ of each state.
We claim that only $N (1 + S)$ bits of classical
communication will be sufficient for such remote preparation,
where $S = - 2 \{ e^2 \log_2 e^2 + ( 0.5 - e^2) [\log_2 ( 0.5 -
e^2)] \} = 1 + H (2e^2) \geq 1 $ [Here, we define
$H(d) = - d \log_2 d - ( 1 -d) \log_2 ( 1-d)$.]
is the entropy of the ensemble.

{\it Remark~F}: Note that the standard teleportation scheme
would require
$2NS$ bits of classical communication. Since $S > 1$
(except for $e^2 = 0 $ or $1/2$),
remote preparation always provides some saving in classical
communication cost over
standard teleportation.

{\bf Proof}: Our proof is analogous to that of Theorem~3.
Let us divide the remote preparation process into two steps.
Starting from
$a_i | 0 \rangle + b_i | 1 \rangle + c_i | 2 \rangle + d_i| 3 \rangle$
in Alice's hand, the goal of the first step is to
obtain an entangled state shared between Alice and Bob,
$a_i | 00 \rangle + b_i | 11 \rangle + c_i | 22 \rangle + d_i| 33
\rangle$.
The goal of the second step is to obtain
$a_i | 0 \rangle + b_i | 1 \rangle + c_i | 2 \rangle + d_i| 3 \rangle$
in Bob's hand.

{\it Step~1}: Start with their $NS$ ebits,
by using entanglement (concentration and) dilution
and their common knowledge of $e$,
Alice and Bob can, with a high fidelity, share $N$ objects of the form
\begin{equation}
\psi = e | 00 \rangle_{AB} + e| 11 \rangle_{AB} + f |22 \rangle_{AB} +
f| 33 \rangle_{AB}
\label{eq:psi}
\end{equation}
where $ f^2 + e^2 = 0.5$, with an asymptotically vanishing
amount of classical communication \cite{LP99,BBPS}.
In what follows, we describe a
procedure that allows Alice and Bob to manipulate
$\psi$ into
$a_i | 00 \rangle_{AB} + b_i | 11 \rangle_{AB} + c_i
| 22 \rangle_{AB} + d_i| 33
\rangle_{AB}$ using only a single classical bit of communication.

First of all, Alice prepares a two-state ancilla in the initial state
$| 0 \rangle_a$. She then couples it with her system $A$
and evolves it with a unitary transformation:
\begin{eqnarray}
e | 0\rangle_a | 0 \rangle_{A} & \to &
(a_i | 0\rangle_a + b_i | 1 \rangle_a ) | 0 \rangle_A \nonumber \\
e | 0\rangle_a | 1 \rangle_{A} & \to &
(b_i | 0\rangle_a + a_i | 1 \rangle_a ) | 1 \rangle_A \nonumber \\
f | 0\rangle_a | 2 \rangle_{A} & \to &
(c_i | 0\rangle_a + d_i | 1 \rangle_a ) | 2 \rangle_A \nonumber \\
f | 0\rangle_a | 3 \rangle_{A} & \to &
(d_i | 0\rangle_a + c_i | 1 \rangle_a ) | 3 \rangle_A
\end{eqnarray}
(It is easy to check that the transformation can be
made unitary. Also, with her knowledge of $a_i$, $b_i$,
$c_i$ and $d_i$,
Alice can, indeed, implement such a unitary transformation.)

Now, starting with
\begin{equation}
| 0 \rangle_a ( e | 00 \rangle_{AB} +
e| 11 \rangle_{AB} + f |22 \rangle_{AB} +
f| 33 \rangle_{AB} ),
\end{equation}
the unitary transformation gives
\begin{eqnarray}
& &|0 \rangle_a ( a_i | 00 \rangle_{AB} + b_i | 11 \rangle_{AB} + c_i
| 22 \rangle_{AB} + d_i| 33
\rangle_{AB} ) \nonumber \\
& + &
|1 \rangle_a ( b_i | 00 \rangle_{AB} + a_i | 11 \rangle_{AB} + d_i
| 22 \rangle_{AB} + c_i| 33
\rangle_{AB} ).
\end{eqnarray}

Now, Alice measures the state of the ancilla, $a$, and
sends the one-bit outcome to Bob. If the outcome is
$0$, the first step of remote preparation is already done.
If the outcome is $1$ instead, Alice and Bob can simply
apply a bi-local unitary transformation to obtain
what is desired. Since each of the $N$ signals requires
one classical bit, $N$ classical bits are sent in the first step.

{\it Step~2}: As in Theorem~3, Alice applies quantum
data compression to the $N$ quantum signals, compressing them
into $NS$ qubits. She can then proceed with the second step of
the remote preparation in the same way as the second step of
the two-stage teleportation, thus sending Bob $NS$ classical
bits.
Adding the classical communication cost in the two steps,
we get $N +NS = N (1+S)$ bits.~QED.

In Theorem~5, the four coefficients $a_i$, $b_i$ and
$c_i$ and $d_i$ are partitioned into two sets with
{\it equal} numbers of elements---$\{a_i, b_i\}$ and $\{c_i, d_i\}$
and the constraint, Eq.~(\ref{eq:constraint}), lies in the
sum of moduli squared of each set.
[Cf. Degeneracy in Schmidt decomposition
of $\psi = e | 00 \rangle_{AB} + e| 11 \rangle_{AB} +
f |22 \rangle_{AB} +
f| 33 \rangle_{AB} $ in
Eq.~(\ref{eq:psi}).] One might wonder if
a partition into sets with equal numbers of
elements is a necessary condition for
reducing classical communication cost. The answer is no,
thanks to the following Lemma.

{\bf Lemma~6}: Suppose that
Alice and Bob share initial entanglement and
a classical communication channel.
Let $| l \rangle$'s be an orthonormal basis of some
Hilbert space, $\cal H$.
Let $I = I_1 \cup I_2 \dots \cup I_M$ be a partition of
the set of indices, i.e., $l$'s.
Suppose Alice would like to help Bob to prepare $N$ objects,
each of which, say the $i$-th one,
\begin{equation}
\psi_i = \sum_l a_{li} | l \rangle
\end{equation}
is a pure state in $\cal H$ and that, for each set $m \in \{1, 2, \cdots,
M\}$, the sum of moduli squared of its elements satisfies
\begin{equation}
\sum_{k \in I_m} | a_{ki}|^2 = c_m
\label{eq:con1}
\end{equation}
for all the states $i$'s in the ensemble.
Here, the sets $I_m$'s and the values $c_m$'s are known to
Alice and Bob in advance while only Alice
knows the individual coefficients $ a_{li}$'s.
We claim that the remote preparation can be done with
$N[(\log_2 d) + S ]$ bits of classical communication,
where $d = l.c.m. \left(|I_1|, | I_2|, \cdots, | I_M|
\right)$ and $S$ is the maximal entropy of the ensemble consisting of
states satisfying
the form Eq.~(\ref{eq:con1}).

{\bf Proof of Lemma~6}: To illustrate the idea of the proof, it
suffices to consider a simple example where $I = I_1 \cup I_2$,
$| I_1| =2$, and $|I_2| =3$. (In this particular example, the
scheme requires a larger amount of classical comunication cost than
direct teleportation and is, therefore, not very useful.)

{\it Step~1}:
By entanglement (concentration and) dilution, Alice and
Bob can manipulate their initially shared entanglement into
$N$ copies of the form
\begin{equation}
\alpha | 00 \rangle + \alpha | 11 \rangle + \beta | 22 \rangle
+ \beta | 33\rangle
+ \beta | 44\rangle
\end{equation}
where $ 2 | \alpha|^2 = c_1$ and $3 |\beta|^2 = c_2$.
For each quantum signal, Alice now prepares an ancilla of dimension
$d= l.c.m. \left(|I_1|, | I_2|, \cdots, | I_M|
\right) $ in the state $| 0 \rangle_a$.
In the current special case, $d=
l.c.m. (2,3) =6$. She now couples the ancilla with each
(say the $i$-th) quantum signal. In the current special case,
she now evolves her combined system of ancilla and the $i$-th quantum
signal
with the following unitary transformation:
\begin{eqnarray}
|0 \rangle_a | 0 \rangle_A
& \to &
\left( | 0 \rangle + | 1 \rangle +
| 2 \rangle
\right)
\left( a_{0i} | 0 \rangle +
a_{1i} | 1 \rangle \right)
| 0 \rangle_A \nonumber \\
|0 \rangle_a | 1 \rangle_A
& \to &
\left( | 0 \rangle + | 1 \rangle +
| 2 \rangle
\right)
\left( a_{1i} | 0 \rangle +
a_{0i} | 1 \rangle \right)
| 1 \rangle_A \nonumber \\
|0 \rangle_a | 2 \rangle_A
& \to &
\left( a_{2i} | 0 \rangle + a_{3i}| 1 \rangle +
a_{4i}| 2 \rangle
\right)
\left( | 0 \rangle + | 1 \rangle \right)
| 2 \rangle_A \nonumber \\
|0 \rangle_a | 3 \rangle_A
& \to &
\left( a_{3i} | 0 \rangle + a_{4i}| 1 \rangle +
a_{2i}| 2 \rangle
\right)
\left( | 0 \rangle + | 1 \rangle  \right)
| 3 \rangle_A \nonumber \\
|0 \rangle_a | 4 \rangle_A
& \to &
\left( a_{4i} | 0 \rangle + a_{2i}| 1 \rangle +
a_{3i}| 2 \rangle
\right)
\left( | 0 \rangle + | 1 \rangle \right)
| 4 \rangle_A .
\label{eq:double}
\end{eqnarray}

Here, the ancilla is further divided into two subsystems,
in the right hand side of the equations. From
the proof of Theorem~5, it is not too hard to see that, by
(i) Alice's measuring the ancilla and sending the outcomes to Bob,
and (ii) Alice and Bob's performing a bi-local unitary transformation,
Alice
and Bob can achieve the first step of remote preparation,
i.e., prepare an entangled state of the form
$a_{0i} | 00 \rangle_{AB} + a_{1i} | 11 \rangle_{AB} +
a_{2i} | 22 \rangle_{AB} + a_{3i} | 22 \rangle_{AB}
+ a_{4i} | 44 \rangle_{AB}$.
For each signal, the first step requires $\log_2 d$ bits of
classical communication.

[Sketch of proof for the general case.
Since, for each $m$, $|I_m|$ divides $d$,
the dimension of the ancilla that Alice has prepared, she
can divide the ancilla into two sub-systems of dimensions $|I_m|$ and
$d \over |I_m|$ respectively. Call them systems $anc^m_1$ and
$anc^m_2$ respectively.
Equivalently, for each $m$,
she can label her ancilla basis vectors by a double index. We
emphasize that, {\it unlike the simple
case presented above}, in the general
case, this double index is a {\it local} labelling depending on
$m$. Denote $|I_m|$ by $R$ and $d \over |I_m| $ by $T$.
Let us use the double index $\{s,t\}$ to
label a basis for the decomposition locally.
Let also $I_m= \{k_1, k_2 , \cdots, k_R\}$.
For $ 1 \leq r \leq R$,
the unitary transformation maps the
initial state
\begin{equation}
|0 \rangle_a | k_r \rangle_A
\end{equation}
into
\begin{equation}
\left( \sum_s a_{{k_{r+s \bmod R~
}}} | s \rangle_{anc^m_1} \right)
\left ( \sum_t { 1 \over \sqrt T}| t \rangle_{anc^m_2} \right)
| k_r \rangle_A ,
\end{equation}
where the ancilla is locally decomposed into two subsystems,
$anc^m_1$ and $anc^m_2$ and its state is labelled by
a double index $(s,t)$.
Suppose Alice measures the ancilla and sends her outcome to Bob. The
outcome can be written, locally for each $m$,
as a pair $s$ and $t$.
$s$ contains the information needed for
the completion of the first step of remote preparation because
it tells Alice and Bob which bi-local unitary transformation to
apply to their states
in the subspace spanned by $\{ | k_r \rangle_A | k_r \in I_m\}$.
On the other hand,
$t$ is unimportant.]

{\it Step~2}: As in the proof of Theorem~5, Alice
applies quantum data compression to her $N$ signals,
compressing them into $NS$ qubits. She then
performs the second step of teleportation, thus
sending $NS$ classical bits to Bob.
By combining the two steps, a total of $N [ (\log_2 d) +S]$ classical bits
are used.~QED.

{\bf Theorem~7}: Suppose that
Alice and Bob share initial entanglement and
a classical communication channel.
Alice would like to help Bob to prepare $N_{tot}$ objects,
each of which, say the $i$-th one,
is of the form $a_i | 0 \rangle + b_i | 1 \rangle +
c e^{i \theta_i} | 2 \rangle$ where
$c$ is known to Alice and Bob in advance,
but $a_i$, $b_i$ and $\theta_i$ are known to Alice only.
We claim that $N_{tot} (S+ 1 - |c|^2 )$ classical bits will be sufficient
for such remote preparation. Here, $S = - [ 2 d^2 \log_2 d^2 +
c^2 \log_2 c^2] $ with $d^2 = ( 1 - c^2 ) /2 $.

{\bf Proof of Theorem~7}:

{\it Idea of the proof}. We set $N_{tot} = N N_1$
where both $N$ and $N_1$ are large and
apply Lemma~6 to
prove Theorem~7.
To do so, it suffices to show that, in the {\it typical}
space of $N_1$ signals, the expression $d$ in Lemma~6
is given by
$ \log_2 d = \log_2
[ l.c.m. \left(|I_1|, | I_2|, \cdots, | I_M| \right) ]
=
N_1 ( 1 - |c|^2) $. The details are as follows.

The Hilbert space of $N_1$ signals is
spanned by the basis vectors $|x_1, x_2, \cdots, x_{N_1} \rangle$.
A normalized basis of the {\it typical}
space (the $| l \rangle$'s in Lemma~6) is
given by vectors of the form
$|x_1, x_2, \cdots, x_{N_1} \rangle$ where between
$N_1 ( |c|^2 - \delta)$ and $N_1 ( |c|^2 + \delta)$
of the $x_i$'s take the value of $2$, for some small $\delta$.
Let us group those $|x_1, x_2, \cdots, x_{N_1} \rangle$
with the same number and locations of $2$'s together.
Within the same group, each of the
$x_i$'s that are not equal to
$2$ can take a value of either $0$ and $1$.
Consequently, there are between $2^{N_1 ( 1- |c|^2 - \delta)}$
and $2^{N_1 ( 1- |c|^2 + \delta)}$ basis vectors in each group.
Furthermore, the weight (i.e., sum of modulus squared of the
wavefunction) of the subspace
corresponding to each group is known in advance to Alice and Bob
as required in Eq.~(\ref{eq:con1}).

The above discussion is rather abstract and can be made clear by
a simple example. Consider the case $N_1 =3$ and $c^2 = 1/2$.
A typical space is spanned by basis vectors $| x_1, x_2, x_3 \rangle$
where one to two of the $x_i$'s take the value of $2$,
i.e., $|2,j_i, j_2 \rangle$, $| j_1, 2, j_2 \rangle$,
$| j_i, j_2, 2 \rangle$ as well as $ | 2, 2, j_1 \rangle$,
$| 2, j_1, 2 \rangle$ and $| j_1, 2, 2 \rangle$ where
$j_i$ takes the value of $0$ or $1$.
Therefore, by fixing the number and locations of the $2$'s,
we have partitioned the typical space into six subspaces.
Furthermore, the weight of each subspace is fixed in advance and
is known to both Alice and Bob. For instance, if
$\psi_i = a_i | 0 \rangle + b_i |1 \rangle + c e^{i\theta_i } | 2
\rangle$, then on expanding $\psi_1 \otimes \psi_2 \otimes \psi_3$,
we find that the projection onto the subspace spanned by
say $\{|2,j_i, j_2 \rangle \}$ is simply
\begin{equation}
c e^{i\theta_1 } | 2 \rangle ( a_2 | 0 \rangle + b_2 |1 \rangle )
( a_3 | 0 \rangle + b_3 |1 \rangle ).
\end{equation}
Its weight is, therefore, simply $c^2 ( 1 - c^2)^2$, independent
of the values of $a_i$ and $b_i$. (Cf. Eq.~(\ref{eq:con1}).)

In summary, the above grouping generally leads to
a partition on the set of basis vectors,
$I = I_1 \cup I_2 \cup \cdots \cup I_M $ where
the size of $I_i$, denoted by $|I_i|$, is
between $2^{N_1 ( 1- |c|^2 - \delta)}$
and $2^{N_1 ( 1- |c|^2 + \delta)}$ and the weight in
each induced subspace satisfies Eq.~(\ref{eq:con1}).
Since $|I_i|$ here is always a power of $2$,
$d= l.c.m.\left( |I_1|, |I_2 |, \cdots, |I_M| \right)$ is
also of order $2^{N_1 ( 1- |c|^2 + \delta)}$.~QED.

\section{Concluding Remarks}
In this paper, we study the classical communication cost in
quantum information processing (CCCIQIP). A motivation of our study is to
better understand the fundamental laws of quantum information processing.
Another motivation is the fact that CCCIQIP can be regarded as a generalization of
quantum communication complexity, a subject of much recent interest.
Our results are as follows.
First, we decompose the usual teleportation
process into a two-step process, a result pointed out by Gottesman.
This leads us immediately to a simple
way to reduce by half the
classical communication cost in entanglement dilution compared
to the earlier scheme \cite{LP99}.
After that, we consider the more general question of
``remote preparation'', a phrase coined by Popescu. Just like
teleportation, Alice and
Bob here share
prior entanglement and also a classical communication
channel. Alice's goal is to help Bob to prepare some state.
Its main difference with usual teleportation
is that we allow Alice to know
exactly the pure state that she is
trying to help Bob to prepare. The question is whether
Alice can somehow reduce the amount of classical communication
using her knowledge on the state.
It is shown here
that, if there are some appropriate constraints on the
ensemble of the states that Alice is trying to send,
Alice will be able to reduce the classical communication
cost below teleportation. We suspect that some constraints
on the ensemble
are necessary for saving classical communication cost. Therefore,
we have the following conjecture.

{\bf Conjecture~8}: (Remote preparation of a general pure state of
a qubit
requires two classical bits of communication?)
Suppose Alice and Bob share prior entanglement
and a classical communication channel only.
Suppose that Alice is asked to help Bob to prepare
$N$ {\it pure} qubit states, $\psi = \psi_1 \otimes \psi_2 \otimes
\dots \otimes \psi_N$
where $\psi_i = a_i | 0 \rangle + b_i | 1 \rangle$ is
an arbitrary pure state of a qubit.
Here, the $a_i$ and $b_i$ are known to Alice but not Bob.
We conjecture that such remote preparation requires $2N$ bits of
classical communication.

{\it Remark~G}: The main difference of the scenario in the above conjecture
from that of the usual teleportation is that here we allow only pure states
but not entangled states.

{\it Remark~H}: A key motivation behind the above conjecture is
the fact that the seemingly inverse process of
the ``randomization'' of an unknown pure state of a single qubit
necessarily generates two classical bits of entropy in the
environment. This result was known to various groups
including a) Sam Braunstein, the author and Tim Spiller
\cite{QIP2000},
b) Alain Tapp and co-workers \cite{Tapp}, c)
Boykin and Roychowdhury \cite{Boykin}
and, d) according to Michael Nielsen \cite{Ben}, proven a few years ago
by Ben Schumacher. Results for the higher dimensional
case have been written up by b), c) and presumably d).

To put things in perspectives, only a few examples of remote preparation have
been studied in this paper. It would, thus, be interesting to consider more
general examples and to attempt to derive a general principle on
the classical communication cost of
remote preparation.
In a more general context, the issue of classical communication cost of
other processes (such as entanglement manipulations \cite{BBPS,LP97},
entanglement purification \cite{purification}) in
quantum information processing deserves careful investigations.
Let us conclude by saying that classical communication cost is only one of the
several types of resources in quantum information processing.
Ultimately, we expect that the fundamental laws of quantum
information processing
will take full accounts of the various types of resources.
The study of the trade-off between qubits and classical communication cost
would be an interesting subject.
Let us conclude by saying that
it is our hope that the study of classical communication cost in
quantum information processing in combination with other
research avenues including \cite{nonlocality,BCT},
will lead us one step closer to the yet
unknown fundamental laws of quantum information processing.

\section{Acknowledgment}
The author particularly thanks D. Gottesman for very helpful discussions
and S. Brauntein for some technical support in the preparation of this
manuscript. Conversations with various
researchers including C. H. Bennett,
S. Braunstein, H. F. Chau, R. de Wolf, D. DiVincenzo, D. Leung, M.
Mosca, M. Nielsen,
S. Popescu, T. Spiller and A. Tapp
are also gratefully acknowledged. Parts of this paper were developed from
work done while the author was at Hewlett-Packard Laboratories, Bristol.


\begin{references}
\bibitem{got} The special case of the result where Alice
knows the state of the qubit---which is all that is
needed for the subsequent sections---was first proven by the author.
During private discussions, Daniel Gottesman pointed out that the result
is, in fact, general.
\end{references}
\end{document}